\documentclass[conference]{IEEEtran}
\IEEEoverridecommandlockouts
\usepackage{cite}
\usepackage{amsmath,amssymb,amsfonts}
\usepackage{algorithmic}
\usepackage{graphicx}
\usepackage{textcomp}
\usepackage{xcolor}
\usepackage{url}
\usepackage{graphicx}
\usepackage{subfigure}
\usepackage{amsfonts,latexsym}
\usepackage{amsmath}
\usepackage{microtype}
\usepackage{booktabs}
\usepackage{subfigure}
\usepackage{tabularx}
\usepackage{multicol}
\usepackage{pdfpages}
\usepackage{multirow}
\usepackage{listings}
\usepackage{xcolor}
\usepackage{xspace}
\usepackage[T1]{fontenc}
\usepackage{caption}

\newcommand{\sysname}{\textsc{{FC-GUARD}}\xspace}
\def\BibTeX{{\rm B\kern-.05em{\sc i\kern-.025em b}\kern-.08em
    T\kern-.1667em\lower.7ex\hbox{E}\kern-.125emX}}
\begin{document}

\title{\sysname: Enabling Anonymous yet Compliant Fiat-to-Cryptocurrency Exchanges}

\author{
Shaoyu~Li\IEEEauthorrefmark{1},
Hexuan~Yu\IEEEauthorrefmark{1},
Md~Mohaimin~Al~Barat\IEEEauthorrefmark{1},
Yang~Xiao\IEEEauthorrefmark{2},
Y. Thomas~Hou\IEEEauthorrefmark{1},
Wenjing~Lou\IEEEauthorrefmark{1}\\
\IEEEauthorblockA{\IEEEauthorrefmark{1}
Virginia Tech, VA, USA\\}
\IEEEauthorblockA{\IEEEauthorrefmark{2}
University of Kentucky, KY, USA\\
}
}

\maketitle

\begin{abstract}
With the rise of decentralized finance, fiat-to-cryptocurrency exchange platforms have become popular entry points into the cryptocurrency ecosystem. However, these platforms frequently fail to ensure adequate privacy protection, as evidenced by real-world breaches that exposed personally identifiable information (PII) and crypto addresses. Such leaks enable adversaries to link real-world identities to cryptocurrency transactions, undermining the presumed anonymity of cryptocurrency use.

We propose \sysname, a privacy-preserving exchange system designed to preserve user anonymity without compromising regulatory compliance in the exchange of fiat currency for cryptocurrencies. Leveraging verifiable credentials and zero-knowledge proof techniques, \sysname enables fiat-to-cryptocurrency exchanges without revealing users' PII or fiat account details. This breaks the linkage between users' real-world identities and their cryptocurrency addresses, thereby upholding anonymity, a fundamental expectation in the cryptocurrency ecosystem. 
In addition, \sysname complies with key regulations over cryptocurrency usage, such as know-your-customer requirements and auditability for tax reporting obligations by integrating a lawful de-anonymization mechanism that allows the auditing authority to identify misbehaving users. This ensures regulatory compliance while defaulting to privacy protection. We implement our system on both desktop and mobile platforms, and our evaluation shows its feasibility for practical deployment.
\end{abstract}


\section{Introduction}
The evolution of blockchain technology has led to the rapid growth of cryptocurrencies such as Bitcoin \cite{nakamoto2008bitcoin} and Ethereum \cite{buterin2013ethereum}, driven by investment demand and decentralized finance (DeFi) applications. In 2025, approximately 861 million people worldwide owned cryptocurrency \cite{lores2025crypto}. For most entrants, the initial step in this ecosystem is the conversion of fiat currencies (e.g., US dollars) into digital assets (e.g., Bitcoin) at market rates. In line with this demand, numerous fiat-to-cryptocurrency exchange platforms \cite{Link,Mercuryo,Moonpay,Robinhood,Sardine} have emerged as primary gateways between traditional financial systems and cryptocurrency markets. 
For example, as of April 2024, reported transaction volumes reached over 2 billion USD for MoonPay and 420 million USD for Transak~\cite{MoonPay_report,Transak_report}.

\noindent\textbf{Privacy risks at cryptocurrency exchange platforms.}
Cryptocurrencies are widely viewed as privacy-preserving alternatives to traditional fiat currencies due to their pseudonymous and decentralized design. However, the operational model of these platforms creates significant privacy risks. A typical fiat-to-cryptocurrency exchange requires the user to transfer fiat currency from their bank account to the exchange platform, which then deposits cryptocurrency into a non-custodial user-controlled cryptocurrency address. To complete this process and comply with regulatory mandates, platforms routinely collect and store sensitive personally identifiable information (PII) and financial data, such as legal names, Social Security Numbers (SSNs), contact details, and bank account numbers. This information is typically co-located with cryptocurrency addresses and transaction logs in centralized databases. Such repositories have repeatedly proven to be high‑value targets for attackers: successive breaches at major exchange platforms \cite{data_breach_1,data_breach_2,data_breach_3,pisces,ledgerdatabreach,Kovacs2020TwitterHackBitcoin} have leaked millions of sensitive records, catalyzing phishing, identity theft, and financial fraud.

\noindent\textbf{Irreversible de-anonymization.} Beyond these direct consequences, a more critical privacy risk arises when adversaries obtain both a user’s PII and the associated cryptocurrency address from compromised exchange records. This correlation connects blockchain pseudonyms to their real‑world identities, nullifying the anonymity that cryptocurrency systems are expected to provide.  Although blockchain pseudonyms do not inherently reveal user identities, transactions on most public blockchains are transparent and permanently accessible. Once the mapping between an address and an identity is established, the pseudonymity of that address is permanently lost, and all past and future transactions become attributable to the identified user \cite{christin2013traveling}. Consequently, even if the complete prevention of data breaches is unrealistic, preventing the establishment of such linkages is critical to maintaining the anonymity properties that underpin cryptocurrency systems.

\begin{figure*}[t]
\centering
    \hspace{-0.27in}
    \subfigure[Current process]{
    \label{subfig:current_design}
    \includegraphics[width=0.40\textwidth]{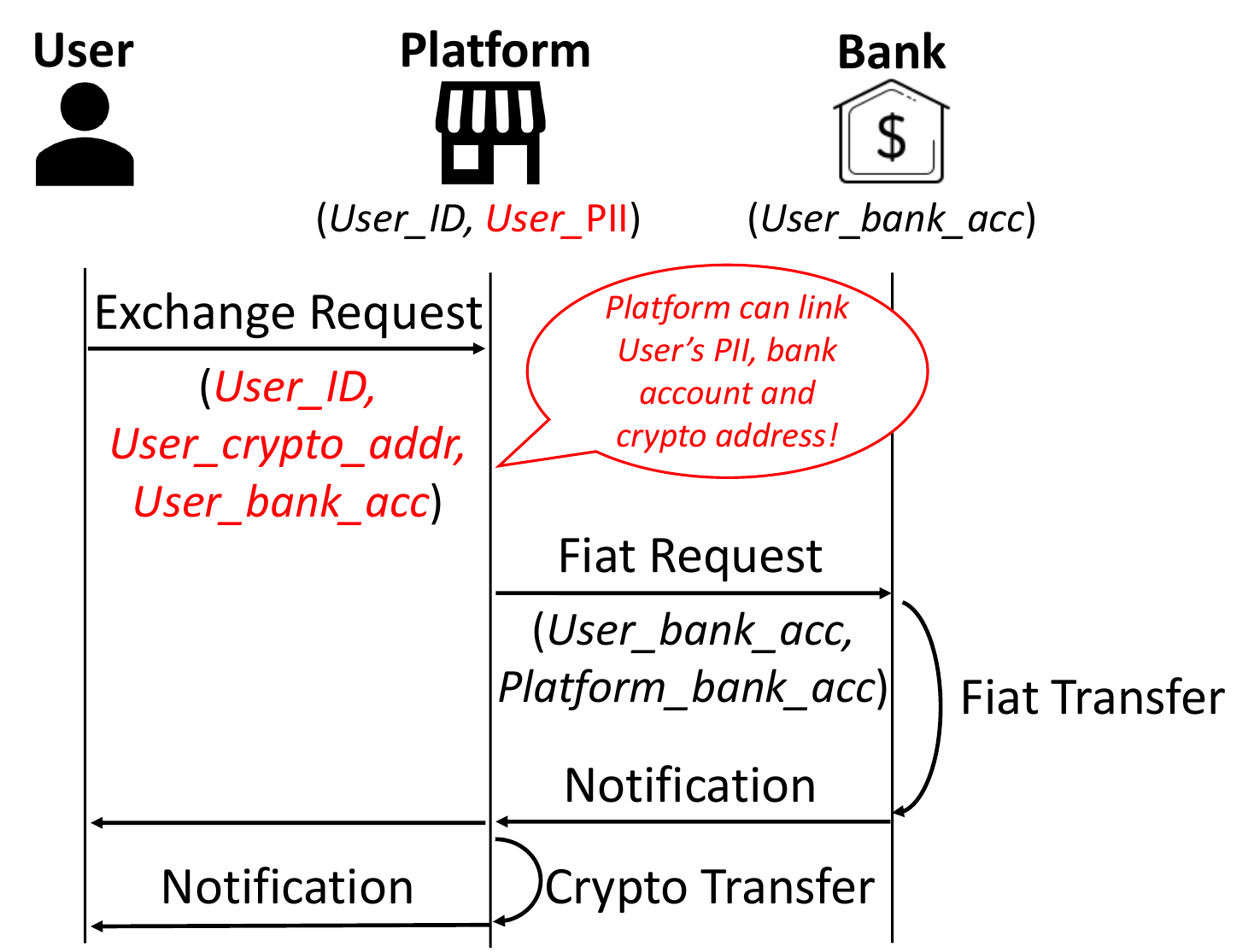}}
    \hspace{-.06in}
    \subfigure[\sysname]{
    \label{subfig:our_design}
    \includegraphics[width=0.42\textwidth]{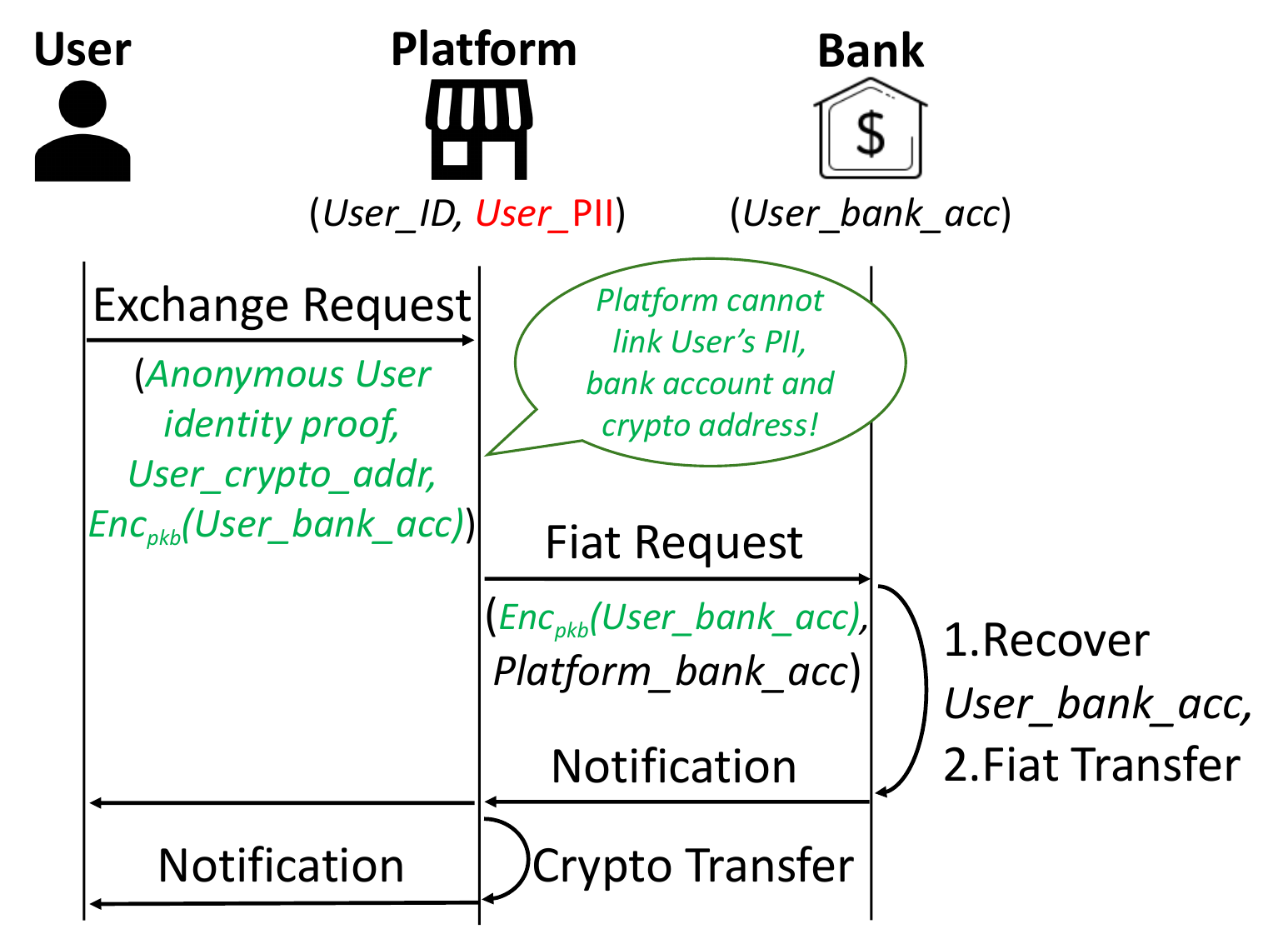}}
    \hspace{-.24in}
    \vspace{-9pt}
\caption{\fontsize{8}{8.75}\selectfont{\textbf{High-level comparison between the current fiat-to-cryptocurrency exchange process and \sysname.} In the current process, users submit their \texttt{User\_ID} (linked to \texttt{User\_PII} in the platform database), \texttt{User\_bank\_acc}, and \texttt{User\_crypto\_addr}, all of which are stored by the platform. This allows adversaries to correlate user identity, bank account, and crypto address. In \sysname, each user instead provides only an anonymous identity proof for authentication, an encrypted bank account for fiat transfer, and a plaintext crypto address. The platform cannot link the user’s identity to the bank or crypto account. Only the bank can decrypt the ciphertext to recover \texttt{User\_bank\_acc} and complete the fiat transfer without learning the crypto address. After the fiat transfer, both designs require the bank to notify both the user and the platform of the transfer status. Once the cryptocurrency is sent, the platform notifies the user to confirm exchange completion.}} \label{fig:The_comprision_of_our_system_and_current_system}
\vspace{-17pt}
\end{figure*}

\noindent \textbf{Regulatory obligations.}
Despite these privacy concerns, exchange platforms must comply with financial regulations in most jurisdictions, including Know Your Customer (KYC) \cite{Cointelegraph-kyc} and tax auditing mandates. KYC requires the collection of PII, such as names, dates of birth, and government-issued identifiers, to verify user identities. In addition, platforms are required to support financial auditing by generating tax reports that reflect users’ gains or losses from cryptocurrency transactions, which are shared with both users and the relevant tax authorities (e.g., the IRS in the US). If a user fails to declare those taxable incomes when filing their tax report, the auditing authority can use these records to identify the user and enforce compliance.

\noindent \textbf{Our goal.}
Our objective is to design a fiat-to-cryptocurrency exchange system that preserves user anonymity by breaking the linkage between users’ fiat currency accounts and their cryptocurrency addresses, while satisfying regulatory requirements. The system should allow an untrusted platform to facilitate exchanges without explicitly linking users' cryptocurrency addresses or transactions to their real-world identities (e.g., PII or bank accounts). As a result, even if the platform is compromised, user privacy remains protected since no explicit mapping between identities and on-chain activity exists.

\noindent \textbf{Regulatory compliance and technical challenges.}
Designing an anonymous yet compliant fiat–to-cryptocurrency exchange introduces several fundamental system and protocol \textbf{challenges}.
First, the platform must support fiat currency transfers without accessing users’ plaintext bank account details. Traditional exchanges require such information to initiate transfers, which creates serious privacy risks. However, enabling secure fiat transfers without disclosing bank account data remains a non-trivial problem. 
Privacy-preserving digital cash like Zcash~\cite{sasson2014zerocash} and Chaumian e-cash\cite{chaum1983blind} emphasize direct peer-to-peer transactions, which differ fundamentally from the conventional user–exchange–bank workflow used by mainstream fiat–to-cryptocurrency exchange platforms.
Second, preserving user anonymity while satisfying regulatory requirements introduces further complexity. KYC procedures require platforms to verify user identities, and tax authorities rely on this information to audit transaction histories. Fully severing the link between a user’s PII and their exchange activity would therefore hinder lawful auditing. While e-cash systems provide strong anonymity, they do not support user-level regulatory accountability. In contrast, fiat–to-cryptocurrency exchanges must enforce KYC and tax reporting, requiring the ability to selectively identify misbehaving users. A practical system must therefore provide privacy by default while still enabling accountable de-anonymization when required.

\noindent \textbf{Our approach.} 
We propose \sysname, a fiat-to-cryptocurrency exchange system that ensures user anonymity and unlinkability while supporting regulatory compliance and compatibility with existing exchange infrastructures. To the best of our knowledge, \sysname is the first system explicitly designed to prevent the linkage between users' real-world identities (e.g., PII and bank accounts) and their cryptocurrency addresses during fiat-to-cryptocurrency exchanges.
\sysname achieves this by leveraging verifiable credentials (VCs) and zero-knowledge proofs (ZKPs) to enable anonymous authentication and transaction processing. The platform can verify user identities and complete fiat transfers \textit{without} learning users’ PII or plaintext bank account information. 
In addition, \sysname includes an \textit{auditing mechanism} where users self-report transaction details by default (consistent with real-world tax reporting practices in the US), and identity de-anonymization is triggered by a trusted auditing authority only when misreporting is detected,  thus preserving privacy for honest users in the common case. 

\sysname is designed to be compatible with existing exchange workflows (see Section~\ref{sub:the_current_design}), emphasizing usability and deployment practicality. A high-level comparison between the current exchange process and \sysname is shown in Fig.~\ref{fig:The_comprision_of_our_system_and_current_system}.
Unlike the current practice, \sysname eliminates the need for users to submit a \textbf{platform-issued user ID} or their \textbf{bank account information} during the exchange, preventing linkage of PII, crypto addresses, and bank details. 
Instead, users rely on VC and ZKP techniques to authenticate and transact anonymously with the platform.
VCs represent users’ PII as credential attributes signed by the platform. 
From a VC, a user can generate unlinkable, one-time verifiable presentations (VPs) that support selective disclosure of attributes. By leveraging ZKPs, a VP allows a user to demonstrate the validity of a pseudonym while only disclosing minimal information necessary for authentication. As a result, when a VP is presented during an exchange, the platform can verify the anonymous user's legitimacy to transact without learning their real-world identity or correlating multiple exchanges initiated by the same user.
To further protect user privacy, \sysname avoids revealing bank account details to the platform. 
Users submit encrypted bank account information under the bank’s public key, along with proofs that the encrypted bank account is bound to the user’s certified identity, allowing fiat transfers to be completed without exposing any bank details to the platform.
Furthermore, to support compliance requirements, we design an auditing mechanism in which users provide encrypted identity information and ZKPs with each transaction. This ensures that identities remain hidden from the exchange, while a trusted auditing authority can selectively recover a user’s identity if misreporting is detected.

\noindent\textbf{Contributions.} In summary, we make the following contributions. 
\textbf{1.} We introduce \sysname, a fiat-to-cryptocurrency exchange system designed to ensure user privacy and anonymity of cryptocurrency addresses using VC and ZKP technologies. \sysname prevents the linkage between users’ real-world identities and cryptocurrency addresses, thereby protecting users' transaction histories from exposure. \textbf{2.} To support regulatory compliance, \sysname enables anonymous yet verifiable KYC based on VCs and ZKPs, and integrates an auditing mechanism to de-anonymize users who misreport their gains or losses from fiat-to-cryptocurrency transactions for tax purposes. \textbf{3.} We implement a \sysname prototype on both desktop and mobile environments and compare it against current systems. Our results show that \sysname adds only minimal computational overhead while significantly enhancing user privacy and anonymity.

\section{Background and Related Work}
\label{sec:background}

\subsection{Current Fiat-to-cryptocurrency Exchange Platform}
\label{sub:the_current_design}

We introduce the registration, exchange, and auditing procedures in traditional fiat-to-cryptocurrency exchange platforms.

\noindent\textbf{User registration.} When accessing the platform for the first time, users must register by submitting PII to comply with KYC regulations. The Social Security Administration (SSA) authenticates each user's identity in advance and issues an SSN for them. The platform utilizes the API from SSA \cite{SSA:eCBSV} to confirm the legitimacy of user identities. Upon successful verification, the platform assigns a unique platform ID to the user, which is linked to the user’s profile and transaction history.

\noindent\textbf{Fiat-to-cryptocurrency exchange.} The exchange process is illustrated in Fig.~\ref{subfig:current_design}. 
To initiate a fiat–to-cryptocurrency exchange, users log in with their platform IDs, and each user submits a request.
After receiving the current exchange rate and confirming it, the user provides the bank account number and cryptocurrency address. The platform then sends the fiat transaction request to its banking partner.
Upon completion of the fiat transfer, the bank returns a receipt to the platform, which subsequently initiates the cryptocurrency transfer from its own wallet to the user's designated address. The platform typically links platform IDs with PII and transaction records, including bank and cryptocurrency details. If compromised, this linkage can enable adversaries to deanonymize users' cryptocurrency identities. Given that most blockchain transactions are publicly visible, adversaries may also infer user behavior.

\noindent\textbf{Financial auditing.}
Exchange platforms record detailed user transactions, including gains and losses, to support tax compliance. These records are compiled into reports provided to both tax authorities and users for personal filings. If a user misreports taxable activity, authorities can use the transaction records and associated identities to trace and enforce compliance.

\subsection{Privacy-preserving Cryptocurrency Systems}

Chaumian e-cash systems \cite{chaum1983blind,dold2019gnu} employ blind signatures to enable anonymous payments, effectively decoupling transaction metadata from user identities. \sysname provides three key advancements over this paradigm: \textit{Identity Verification.}
Chaumian e-cash lacks built-in identity verification, requiring separate KYC processes. In contrast, \sysname enables selective identity disclosure and verification through VCs and ZKPs. \textit{User Auditing Capability.}
Chaumian e-cash does not support user-level regulatory auditing. \sysname enables authorities to audit reported transactions and deanonymize misbehaving users based on VCs and ZKPs.
\textit{Compatibility.}
Chaumian systems follow a payment architecture that differs from conventional user–exchange–bank workflows \cite{Moonpay,Robinhood,Sardine}, requiring substantial changes to existing infrastructure. \sysname instead preserves compatibility with current exchange payment flows, reducing deployment barriers.

In parallel, numerous efforts have focused on preserving privacy in intra-cryptocurrency payments.
Zcash \cite{sasson2014zerocash} is a standalone cryptocurrency that uses a specific type of ZKP (zk-SNARKs) to ensure the anonymity of its users and their transactions on its blockchain. Buterin et al. \cite{buterin2024blockchain} extended this technology and proposed the privacy-pool protocol.
\textit{Zether} and \textit{Anonymous Zether} \cite{bunz2020zether,diamond2021many} apply ZKPs to provide confidential payments over Ethereum smart contracts.
Solutions like \cite{gjosteen2022pribank,green2017bolt,heilman2017tumblebit,ng2021ldsp,qin2023blindhub} focus on providing privacy for off-chain payment methods. 
\textit{Bolt} \cite{green2017bolt} constructs anonymous payment channels built on micropayment channels and utilizes the combination of blind signature and ZKP to protect user anonymity in transactions. \textit{LDSP} \cite{ng2021ldsp} and \textit{PriBank} \cite{gjosteen2022pribank} both address the privacy issues in layer-2 solutions for cryptocurrency payments, with LDSP focusing on payer privacy in a unidirectional setup \cite{riahi2024bitcoin}, and PriBank enhancing private balances and transaction values by using ZKP. \textit{TumbleBit} \cite{heilman2017tumblebit} achieves unlinkability by leveraging secure multi-party computation and ZKP. Additionally, methods such as coin mixing, proposed by \cite{glaeser2022foundations,le2021amr}, enable users to obscure their cryptocurrency transactions, thereby facilitating unlinkable payments and preventing both the service provider and users from tracking coins. 
However, the effectiveness of coin mixing depends on pool size and usage: when participation is limited, adversaries can link inputs to outputs, reducing privacy. In contrast, \sysname achieves per-transaction anonymity via VCs and ZKPs, eliminating the need for mixing pools.
\textit{Pisces} \cite{pisces} supports anonymous and compliant crypto-to-crypto exchanges but does not apply to fiat-to-cryptocurrency settings because it does not conceal users' bank accounts from the platform.
Recent works on privacy-preserving central bank digital currencies (CBDCs), such as Platypus\cite{wust2022platypus} and Peredi \cite{kiayias2022peredi}, focus on crypto-native fiat issued by central banks. While they provide anonymity and regulatory compliance, they do not address the exchange of account-based fiat in commercial banks for decentralized cryptocurrencies, which is the focus of \sysname.
In summary, prior works primarily focus on privacy within crypto-native systems, overlooking fiat-to-cryptocurrency exchanges where user anonymity is most vulnerable. 

\section{System Overview}
\label{sec:system_overview}

\begin{figure*}[t]
    \centering
    \includegraphics[width=0.75\linewidth]{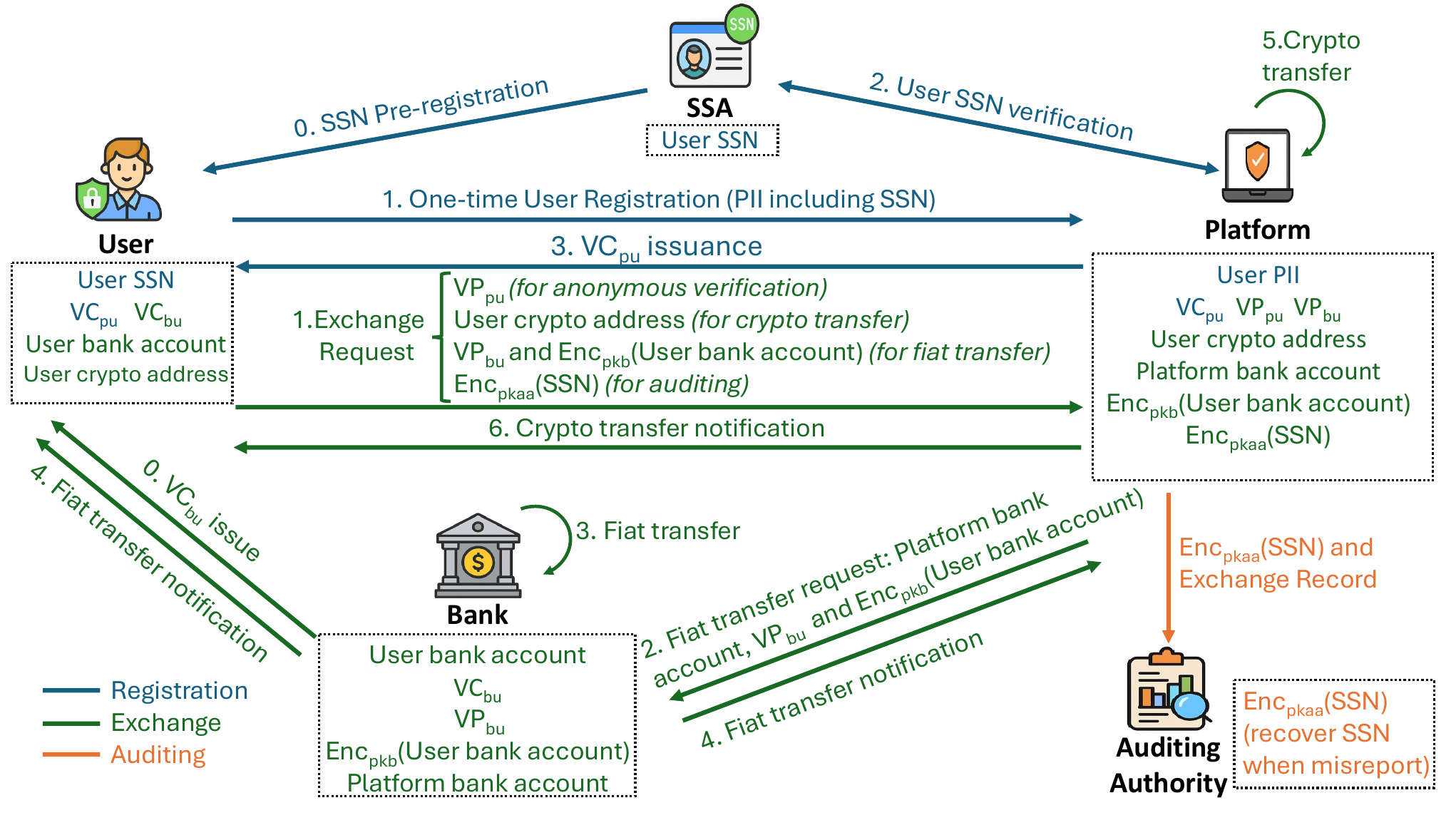}
    \vspace{-10pt}
    \caption{\fontsize{8}{8.75}\selectfont{\textbf{\sysname system architecture.} The workflow of \sysname consists of three main phases: \textbf{registration}, \textbf{exchange}, and \textbf{auditing}. 
Each entity is shown with a surrounding box that lists the information it can \textit{access} and \textit{process} during protocol execution. This design ensures that no single entity can link a user's real identity to their cryptocurrency transactions.
}}
    \label{fig:system architecture}
    \vspace{-20pt}
\end{figure*}

\subsection{System Stakeholders and Threat Model}

\sysname involves five entities: the exchange platform, users, bank, Social Security Administration (SSA), and auditing authority. 
\textbf{Exchange Platform} acts as the central entity facilitating the fiat-to-cryptocurrency exchange process and serves three primary functions. 
First, it handles user registration and identity verification in compliance with KYC regulations. 
Second, it executes fiat-to-cryptocurrency orders by interacting with user-provided crypto addresses and the banking partner. 
Third, it fulfills financial regulatory obligations by compiling and submitting transaction records to the auditing authority.
\textbf{Users} utilize the platform to exchange fiat currency and cryptocurrency anonymously, using their self-controlled, non-custodial cryptocurrency addresses. 
In compliance with KYC and other regulatory requirements, users must maintain valid identities (e.g., SSNs issued by SSA) and sufficient bank account balances to conduct transactions. Invalid identities or insufficient funds will lead to transaction failures.
Additionally, users must adhere to financial regulations such as tax reporting procedures. Users who misreport gains or losses to tax authorities are subject to de-anonymization. In contrast, the information provided by compliant users is inherently verifiable, eliminating the need for deanonymization.
\textbf{Bank} aims to facilitate fiat currency transfers between user accounts and platform accounts. \textbf{SSA} refers to the governmental entity responsible for issuing and verifying legitimate identities (e.g., SSN), which may vary across jurisdictions. In the US, the SSA provides APIs (e.g., \texttt{eCBSV}~\cite{SSA:eCBSV}) for service providers to verify user identities, and \sysname also adheres to this process. 
\textbf{Auditing Authority} monitors users’ transaction gains and losses from the platform reports. \sysname focuses primarily on tax auditing, where an authority (e.g., IRS) can track every transaction, aligning with real-world scenarios.

\sysname assumes that the \textbf{platform} and the \textbf{bank} are \textit{honest-but-curious}, meaning that while they may attempt to link a user's cryptocurrency address to its PII during a particular exchange, they follow the prescribed protocols. This includes performing accurate fiat and cryptocurrency transfers and ensuring transaction atomicity without premature termination. The platform does not access users’ PII or plaintext bank account numbers during exchange transactions. The bank, on the other hand, knows users’ real-world identities and fiat account details and is aware of fiat transactions between users and the platform. However, it does not have access to users’ cryptocurrency addresses or detailed exchange information.
This separation ensures that no single entity can independently associate a user’s real identity with their cryptocurrency activities.
The \textbf{SSA} and \textbf{auditing authority} are considered \textit{trusted}. The SSA provides SSN verification services only during user registration with the platform, while the auditing authority can de-anonymize misbehaving users during the auditing phase. Neither of the two entities can obtain users’ cryptocurrency addresses. In addition, scenarios involving collusion between the platform and the bank are excluded.
We assume all communication channels between parties are secured using standard TLS mechanisms. Following existing platforms, deanonymization of the network layer or the software layer is beyond the scope of this work \cite{pisces}.

\subsection{System Architecture and Security Goals}

\sysname enables users to purchase cryptocurrencies with fiat currency anonymously by preserving the overall workflow of conventional exchanges while replacing direct disclosures of personal and financial information with privacy-preserving computations. 
Specifically, \sysname leverages VC and ZKP techniques to verify the validity of anonymous users' identities and ensure their specific attributes meet predefined criteria. This allows the platform to confirm a user’s eligibility for exchange without accessing any plaintext PII or bank account details.
To achieve unlinkability and confidentiality, \sysname applies probabilistic public-key encryption to sensitive fields such as SSNs and bank account numbers. The platform validates user-submitted ZKPs that prove the correctness of encrypted attributes without learning their actual values.
Only designated parties (the bank for account transfers and the auditing authority for compliance checks) can decrypt the corresponding ciphertexts when necessary. This design ensures that no single entity can associate a user's real identity with their cryptocurrency transactions.

\noindent \textbf{\sysname consists of three phases:}

\noindent \textbf{1. One-time user registration (non-anonymous).} 
The user submits their PII (including SSN) to the platform, which verifies it with the SSA through the SSA’s API. Upon successful verification, the platform issues a verifiable credential \(VC_{pu}\) to the user, certifying that their SSN is valid.

\noindent \textbf{2. Fiat-to-cryptocurrency exchange (anonymous).} 
\sysname ensures user privacy by decoupling the user registration with the fiat-to-cryptocurrency exchange process. When the user intends to anonymously exchange fiat currency for cryptocurrency on the platform, it generates a \(VP_{pu}\) derived from the \(VC_{pu}\) it obtained during registration. This \(VP_{pu}\) can be verified by the platform, which cannot link the \(VP_{pu}\) to the original recorded \(VC_{pu}\). After identity verification, the user provides the platform with another \(VP_{bu}\), derived from a \(VC_{bu}\) issued by the bank. This \(VP_{bu}\) only reveals the attribute relevant to the bank’s name to the platform. Additionally, the user prepares an encryption of its bank account under the bank’s public key. The platform then relays this \(VP_{bu}\) and ciphertext to the designated bank. The bank verifies the validity of the \(VP_{bu}\) and ZKP, decrypts the ciphertext to obtain the user’s bank account number, and can use multi-factor authentication methods, such as sending a one-time SMS code to the user, to prevent fraudulent transactions. Once verified, the bank sends the fiat currency to the platform, which then proceeds with the cryptocurrency transaction.

\noindent \textbf{3. Auditing process.} 
For each transaction, the user encrypts its SSN under the auditing authority’s public key as $\mathsf{Enc}_{pk_{aa}}$(SSN), and provides a ZKP proving that this encrypted SSN matches the hidden SSN attribute in \(VP_{pu}\).
The platform verifies this proof before processing the order and submits the encrypted SSN along with transaction metadata to the auditing authority. 
By default, user identities remain hidden. However, if the user misreports taxable gains or losses, the authority may decrypt $\mathsf{Enc}_{pk_{aa}}$(SSN) to recover the user's identity and enforce tax compliance.
Although the auditing authority holds the cryptographic ability to deanonymize all transactions, our design ensures this power is only exercised when a user fails to self-report, preserving user privacy by default.

\sysname eliminates the need to disclose PII and bank account information during fiat-to-cryptocurrency transactions. However, the platform can still verify user identity by validating the VP and ZKP, ensuring it knows a cryptocurrency address without linking it to the user’s real identity. Nevertheless, banks and auditing authorities can identify the user based on the identity encrypted under their public keys.

\noindent\textbf{The security goals of \sysname are as follows:
}

\noindent
\textbf{Anonymity.}
Adversaries should not be able to identify the user behind a particular cryptocurrency transaction or cryptocurrency address. Specifically, they cannot link the user’s cryptocurrency address used during the fiat-to-cryptocurrency exchange process with the PII recorded during registration.


\noindent \textbf{Unlinkability.}
Given multiple fiat-to-cryptocurrency exchange orders placed under different pseudonyms, adversaries cannot link pseudonyms to the same user.

\noindent \textbf{Data confidentiality.}
For a specific user, adversaries should not obtain its bank account information or exchange activities from the platform's leaked data.

\noindent \textbf{Accountability.} Both users and the platform must adhere to regulatory guidelines. Users must accurately report transactions, while the platform relays all required data to auditing authorities without modification. The integrity of this process is crucial to prevent potential deception, as long as there is no collusion between users and the platform. In our design, accountability mechanisms are critical, particularly when users fail to report accurately. The platform assists auditing authorities in correctly identifying users and their transactions, ensuring compliance and integrity in auditing.
\section{\sysname Detailed Design}
\label{sec:protocol}

\subsection{Preliminaries}
\label{subsec:prelim}

\noindent\textbf{Verifiable Credentials (VCs)}, also known as anonymous credentials (ACs)
\cite{camenisch2003signature}, are constructed using cryptographic techniques such as blind signatures (e.g., CL signatures~\cite{camenisch2003signature}, BBS+ signatures~\cite{boneh2004short}), and ZKP. VCs provide strong security and privacy guarantees through selective disclosure, allowing users to prove that they meet certain criteria by revealing only necessary attributes, thereby enhancing privacy and control over personal data. 

A VC system involves three entities: the issuer, holder, and verifier. The issuer creates and signs credentials attesting to a holder’s attributes (e.g., identity). The holder stores these credentials in secure software \cite{w3VerifiableCredentials} and can generate a one-time unlinkable VP by randomizing the issuer’s signature. With ZKPs, the holder proves to the verifier that the VP is derived from a valid VC and satisfies requested attribute constraints, without revealing sensitive data. This ensures anonymity and unlinkability, even when the verifier is the original issuer.

\begin{figure}[t]
\centering
    \hspace{-.1in}
    \subfigure[Credential issuance]{
    \label{subfig:Credential_issue}
    \includegraphics[width=0.26\textwidth]{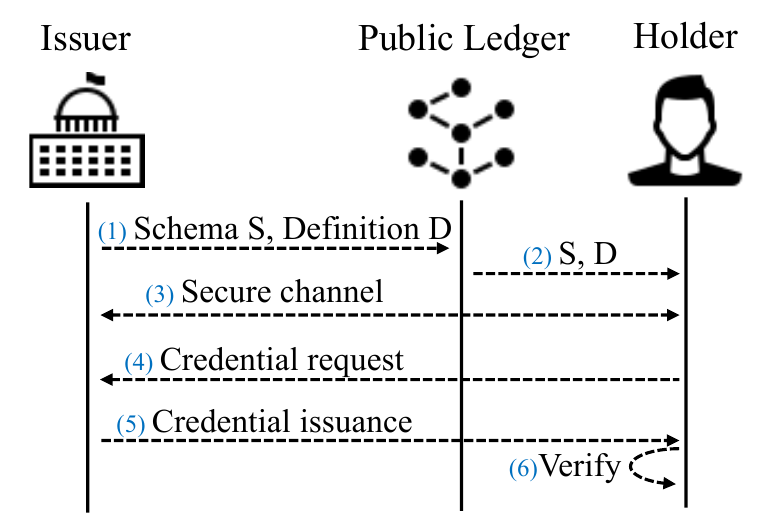}}
    \hspace{-.1in}
    \subfigure[Presentation verification]{
    \label{subfig:Credential_Verify}
    \includegraphics[width=0.204\textwidth]{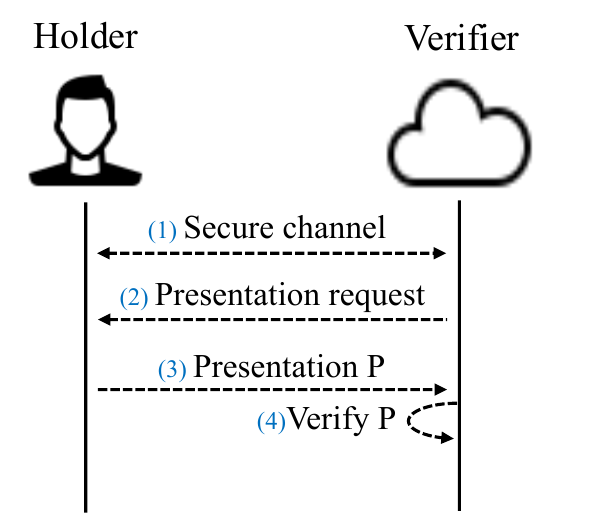}}
    \vspace{-8pt}
\caption{Credential issuance and verification protocols} \label{fig:Credential_protocol}
\vspace{-18pt}
\end{figure}

The credential issuance protocol is illustrated in Fig.~\ref{subfig:Credential_issue}. In message 1, the issuer defines a credential schema \(S\) and a credential definition \(D\) and publishes them in the public ledger. \sysname defines two types of schema issuers: platform and bank. For the platform, the schema attributes are \texttt{[name, birthday, SSN]}. For the Bank, the attributes are \texttt{[bank's name, bank account, SSN]}. The credential definition includes the schema ID and issuer's public key \(pk\), used for verifying CL signatures. To generate \(pk\), the issuer first produces two random 1536-bit primes $p', q'$ such that $p \leftarrow 2p' + 1$ and $q \leftarrow 2q' + 1$ are primes too. After computing $n \leftarrow pq$, the issuer generates a random quadratic residue \( S \) modulo \( n \) and random $x_Z, x_{R_1}, \ldots, x_{R_l} \in [2, p'q' - 1]$. Then issuer computes \(Z \leftarrow S^{x_Z} \pmod{n}\);\( \quad \{R_i \leftarrow S^{x_{R_i}} \pmod{n}\}_{1 \leq i \leq l}\), defining issuer’s public key as $p_k = (n, S, Z, \{R_i\}_{1 \leq i \leq l})$ and the private key as $s_k = (p, q)$. In message 2, the holder retrieves \(S\) and \(D\) from the public ledger and obtains \(p_k\) in \(D\). Step 3 involves mutual authentication between the issuer and holder to set up a secure channel. 
The holder creates a private \textit{link secret} to link multiple credentials. This secret is blinded by the holder, who also produces a corresponding proof. Both the blinded \textit{link secret} and its proof are sent to the issuer in the credential request. The issuer will verify the credential request and generate a credential \(C\) for the holder. In \(C\), the issuer encodes all attributes to integers and uses its private key \(s_k\) to sign the credential. \(C\) includes the signature and the proof of the signature, which can be verified by the holder when receiving \(C\). The verification is shown in Fig.~\ref{subfig:Credential_Verify}. Upon request from a verifier, the holder constructs a VP. VP selectively obscures attributes that the holder opts not to disclose, thereby preserving privacy while maintaining the integrity of the credential's information. The verifier can verify the VP's signature using the issuer's public key. Furthermore, based on the verifier's request, the issuer can prove that certain concealed attributes meet specified requirements without revealing the attributes themselves (i.e., selective disclosure). 

VC technique is a key enabler for Decentralized Identifiers (DIDs) standardized by the World Wide Web Consortium (W3C) \cite{W3CDecentralizedIdentifiers}. 
DIDs represent verifiable, self-owned digital identities and are increasingly adopted in both public and private sectors~\cite{ukundefinedDigitalIdentity}. Each DID resolves to a document containing public keys stored in a verifiable registry (e.g., distributed ledger).
Although numerous frameworks for VCs (or ACs) exist \cite{yu2024aaka,du2023ucblocker}, \sysname adheres to the VC scheme standardized by the W3C~\cite{w3VerifiableCredentials} and Internet Engineering Task Force (IETF)~\cite{ietfSDJWTbasedVerifiable}, ensuring compatibility and interoperability of \sysname across standard web applications.

\noindent\textbf{Zero-Knowledge Proofs (ZKPs)}~\cite{goldreich1994definitions} enable a prover to demonstrate knowledge of a secret without disclosing it to a verifier. ZKP is frequently used alongside VC, augmenting its utility in securing digital identities and various applications by providing robust proof mechanisms without compromising data confidentiality. 
The ZKP mechanism implemented in \sysname is based on non-interactive zero-knowledge (NIZK) protocols, specifically utilizing BBS+ signature-based proofs.

\subsection{Initialization and User Registration Process}

We require a user \(U\) to pre-register with \(SSA\) and keep SSA-authenticated PII: \(PII_u = (name, birthday, SSN_{u})\). \(SSA\) provides an \(API_{SSA}\) for platforms to verify \(U\)'s identity, as is common for general financial applications. 
The bank \(B\) and auditing authority \(AA\) each hold a public-private key pairs, \((pk_{\text{b}}, sk_{\text{b}})\), and \((pk_{\text{aa}}, sk_{\text{aa}})\), respectively.  
As a regular bank customer, \(U\) is pre-issued a credential \(VC_{bu}\) from \(B\):
\( VC_{bu} = (name_{b}, AccNum_{bu}, SSN_u, \sigma_{bu}) \), 
where \(name_b\) is the bank's name, \(AccNum_{bu}\) is the user's bank account number, \(SSN_u\) is the user's SSN, and \(\sigma_{bu}\) is the credential signature.

\noindent\textbf{User registration.}
To register, the exchange platform \(P\) must verify the identity of \(U\) (as shown in Fig.~\ref{fig:User_Registration_Process}). 
First, \(U\) sends personally identifiable information \(PII_u\) to \(P\), which then invokes \(API_{SSA}\) to verify \(PII_u\). Upon successful verification, \(P\) issues \(U\) a verifiable credential \( VC_{pu} = (name, birthday, SSN_u, \sigma_{pu}) \), where \(\sigma_{pu}\) is the credential signature. \(U\) validates \(VC_{pu}\) using $P$'s public key and stores it in a confidential wallet.

\subsection{Fiat-to-cryptocurrency Exchange Process}
\label{Fiat-cryptocurrency Exchange Process}
\begin{figure}[t]
\centering
\includegraphics[width=0.9\linewidth]{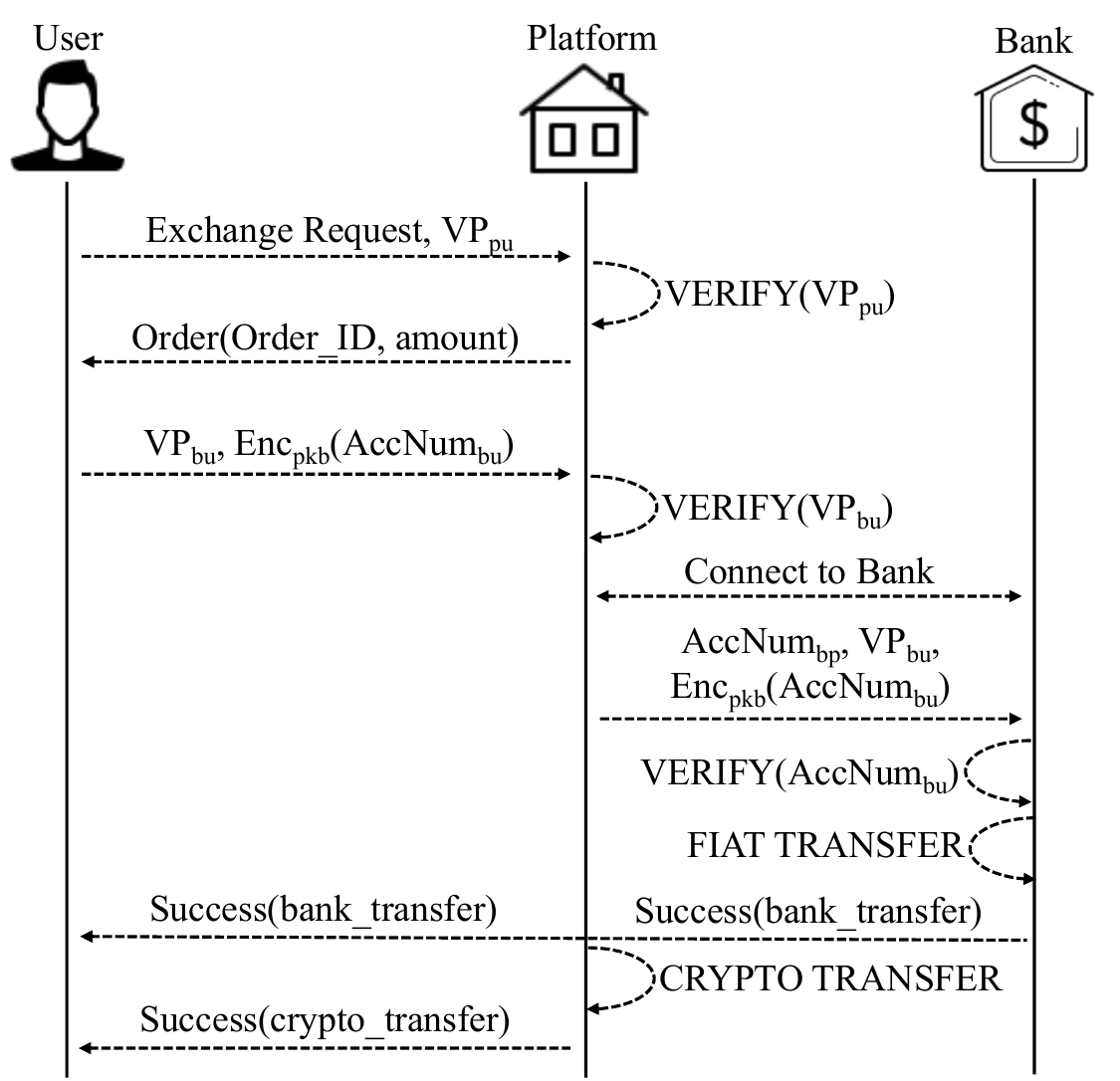}
\caption{Fiat-to-cryptocurrency exchange process}
\label{fig:Exchange_Process}
\vspace{-20pt}
\end{figure}

\noindent The exchange process illustrated in Fig.~\ref{fig:Exchange_Process} includes three steps:

\begin{figure}[t]
\centering
\includegraphics[width=0.9\linewidth]{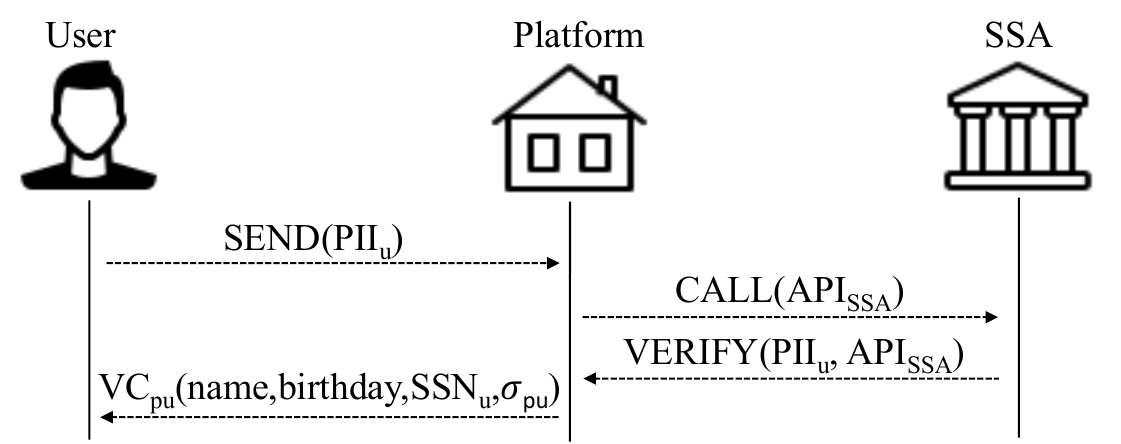}
\vspace{-2pt}
\caption{User registration process}
\label{fig:User_Registration_Process}
\vspace{-17pt}
\end{figure}

 \noindent \underline{\textbf{Step-1: Platform User Verification.~}}
The initial step in the exchange process involves verifying the user's identity and reaching an order agreement.
When exchanging fiat for cryptocurrency, \(U\) initiates an exchange request on \(P\), specifying the desired cryptocurrency type, quantity, and its cryptocurrency address \(CryptoAddr_u\). 
To verify the identity of \(P\), \(U\) generates a VP based on the VC issued by the platform during the registration process. The VP can be represented as:
\( VP_{pu}=(name', birthday', SSN_u', Proof_{pu}) \),
where \(attribute'\) denotes the value of this \(attribute\) which will be concealed from \(P\); $Proof_{pu}$ is a ZKP proving the validity of $U$ as the platform's user, but without revealing which specific user.
If \(P\) has additional requirements regarding user identity, such as specifying that the user's age must be more than 
\(x\), $Proof_{pu}$ can include a range proof
based on the attribute \(birthday\) showing the binary fact that the age is larger than $x$. Then, $U$ sends $VP_{pu}$ to \(P\).
After the verification of \(VP_{pu}\) through $Proof_{pu}$, \(P\) generates an order containing an order ID and the amount of fiat currency needed, and then sends it to \(U\).

\noindent \underline{\textbf{Step-2: Bank User Verification.~}}
This step is a crucial prerequisite of ensuring currency transfer as it allows the platform $P$ and the bank $B$ to verify the user $U$'s ownership of a valid bank account.
To maintain current business logic and ensure user convenience, our design includes the user submitting its anonymous bank account information to the platform. Upon verifying the correctness of the user's bank information, the platform bundles its own bank account details with the user's bank information and submits them to the bank. The bank then verifies the user's identity.
In detail, \(U\) first generates the ciphertext \(Enc_{pkb}(AccNum_{bu})\) by encrypting $AccNum_{bu}$ with \(B\)'s public key  \(pk_{\text{b}}\). Then 
\(U\) generates a verifiable presentation \(VP_{bu}\) from \(VC_{bu}\) as follows: 
\( VP_{bu}=(name_b, AccNum_{bu}', SSN_u', Proof_{bu}) \),
where the values of the user's account number and SSN are concealed; only the bank name is revealed. 
$Proof_{bu}$ is a ZKP that shows the user bank account number ciphertext \(Enc_{pkb}(AccNum_{bu})\) is correctly generated as well as the validity of $name_b$. 
Furthermore, the user generates an additional proof \( Proof_{eq}\left( SSN_u' \, (\text{in}\ VP_{pu}) = SSN_u' \, (\text{in}\ VP_{bu}) \right) \) to demonstrate the consistency of the hidden $SSN$ attributes in \( VP_{pu} \) and \( VP_{bu} \) without disclosing the value.
$U$ sends $VP_{bu}$ and \(Proof_{eq}\) to $P$ for verification. 
Then $P$ retrieves \(name_b\) from $VP_{bu}$ and connects to the bank $B$.
After connecting, \(P\) sends the message \(M_{pb} = (AccNum_{bp},VP_{bu},Enc_{pkb}(AccNum_{bu}))\) to \(B\). \(B\) uses its private key \( sk_{\text{b}} \) to get \(U\)'s bank account number \(AccNum_{bu} =  Dec_{skb}(Enc_{pkb}(AccNum_{bu})) \) and verify the correctness of \(AccNum_{bu}\) based on \(Proof_{bu}\). 

\noindent \underline{\textbf{Step-3: Fiat-to-cryptocurrency Transfer.~}}
Upon verifying $U$, bank $B$ performs the fiat currency transaction to transfer the required amount from \(AccNum_{bu}\) to \(AccNum_{bp} \) and generate a transaction receipt. 
After the successful fiat currency transfer, \(B\) will send the transaction receipt separately to both \(P\) and \(U\), typically via SMS. \(P\) proceeds with the cryptocurrency transfer by transferring the agreed-upon amount of cryptocurrency from its crypto address \(CryptoAddr_p\) to \(U\)'s crypto address \(CryptoAddr_u\). Because cryptocurrency transactions are public on the decentralized ledger, \(B\) can infer \(CryptoAddr_u\) by observing the platform's crypto transaction recipient on the ledger. 
Therefore, \(P\) implements two obfuscation methods: First, \(P\) maintains a pool of crypto addresses and requires 
\(U\) to provide multiple addresses, distributing transactions across distinct platform-user address pairs. \(P\) enforces periodic address rotation to break long-term linkage. Second, \(P\) adds random delays for both fiat-to-cryptocurrency conversions and multi-address transaction distribution.
In addition, \(U\) should use a different \(CryptoAddr_u\) for each order \cite{bitcoin_privacy} to ensure unlinkability.

We remark that in \textbf{Step-2}, to ensure that encrypted ciphertexts appear randomized across different transactions, thereby preventing the linkage of transactions to a single user, users should employ probabilistic encryption schemes such as ElGamal~\cite{elgamal1985public} and Paillier~\cite{paillier1999public} cryptosystems. These ensure that encrypting the same plaintext multiple times results in different ciphertexts each time due to the use of random values in the encryption process, ensuring unlinkability.

Crucially, in the entire process, the platform is not aware of the user's identity or bank account. Verification of sufficient funds is handled internally by the bank. Hence, our system avoids the risk of financial data exposure in existing exchange practices, where the platform needs to collect user bank account details before submitting a debiting transaction to the bank.

\subsection{Auditing}

Auditing authorities need to monitor users' financial activities, but due to the anonymity of user identities, platforms cannot directly link financial activities to real identities. 
Our protocol is designed to identify users associated with specific unreported transactions through an optimized exchange mechanism, as opposed to tracking a user's complete transaction history. 
In the user's identity verification step, \(U\) needs to encrypt its \(SSN_u\) using auditing authority \(AA\)'s public key \(pk_{\text{aa}} \), denoted as \(Enc_{pkaa}(SSN_u)\). After \(U\) generates \(VP_{pu}(name', birthday', SSN_u')\), \(U\) has to generate an extra proof \(Proof_{ua}(Enc_{pkaa}(SSN_u')=Enc_{pkaa}(SSN_u))\) and send them together to \(P\). \(P\) can verify the proof and if  \(Verify(Proof_{ua}) = True
\), \(P\) will store the \(Enc_{pkaa}(SSN_u)\) in the exchange record. As discussed in Section \ref{Fiat-cryptocurrency Exchange Process}, \(U\) should employ probabilistic encryption schemes to ensure that the ciphertext \(Enc_{pkaa}(SSN_u)\) exhibits randomness properties.

After each exchange, \(P\) will hide \(U\)'s \(CryptoAddr_u\) and the details of the crypto transfer in the exchange record and send the record to \(AA\). If \(AA\) receives the auditing report about this exchange from \(U\), they only need to check the fiat transfer amounts and other necessary information. However, if \(AA\) does not receive \(U\)'s report in time, \(AA\) uses its private key \(sk_{\text{aa}}\) to decrypt \(U\)'s SSN: \(SSN_u = Dec_{skaa}(Enc_{pkaa}(SSN_u))\), which can help \(AA\) identify \(U\).

\subsection{Security Analysis}
\label{sec:analysis}

\sysname aims to protect user anonymity and privacy in exchanges between fiat and cryptocurrency by employing VC, ZKP, and public key cryptosystems. The following analysis addresses anonymity, unlinkability, correctness, communication security, accountability, and replay attack protection.

\noindent\textbf{Anonymity and crypto unlinkability.} 
\sysname ensures anonymity by utilizing the VC and ZKP schemes in the registration and exchange process. During the VP process, users conceal attributes potentially linking to PII, and utilize ZKPs to prove necessary conditions. The adversary cannot link the observed user's PII during the registration process and the cryptocurrency address during the exchange process. In addition, in different exchange processes, users will employ distinct cryptocurrency addresses for anonymity protection. Users generate different VPs based on the same VC, which ensures that all of a user's cryptocurrency addresses cannot be linked, thereby enhancing the anonymity and privacy of user cryptocurrency identities.

\noindent\textbf{Data confidentiality.}
\sysname enhances user privacy by collecting the minimum amount of user information, such as bank accounts. Through the use of VP and ZKP schemes, users ensure that the platform never directly accesses their bank account information during the bank transfer. Furthermore, each exchange record is decoupled from the user’s PII, ensuring that adversaries cannot link an exchange activity to a specific user, thereby safeguarding user privacy.

\noindent\textbf{Exchange unlinkability.} \sysname ensures that a user’s activities across different exchanges cannot be linked.  This is achieved by introducing randomness in each VP and ZKP generation. Even when derived from the same VC and proof request, outputs appears distinct. Likewise, the ciphertexts produced by encrypting the same information using our adopted probabilistic asymmetric encryption schemes also undergoes randomization for different exchanges. To further reduce linkability, users should use a different cryptocurrency address for every transaction, a widely recommended practice in the cryptocurrency community \cite{bitcoin_privacy}.

\noindent\textbf{Correctness.} 
All entities are assumed to follow the prescribed protocols. The correctness of \sysname depends on the accuracy of the well-established cryptographic building blocks it utilizes, including VCs, ZKPs, and public key cryptosystems (e.g., ElGamal or Paillier).

\noindent\textbf{Communication security.}
\sysname is secure against outsiders, as it utilizes secure transmission links by validating network message integrity through the TCP protocol and encrypting channels with TLS to prevent eavesdropping and tampering. This ensures the integrity and confidentiality of communication content in \sysname and upholds the assumption of secure communication.

\noindent\textbf{Accountability (auditability).}
\sysname ensures that auditing authorities can effectively monitor every user transaction. If transactions are not correctly reported, authorities can de-anonymize misbehaving users using their own secret keys, while keeping the platform unaware of the specific user identities. The randomness introduced by the probabilistic encryption scheme ensures that each ciphertext (the encrypted identities) appears uniformly distributed to the untrusted platform, thus preserving user privacy.

\noindent\textbf{Replay attack protection.}
To prevent adversaries from attempting to reuse a victim's VPs, ZKPs, or ciphertexts, multi-factor authentication can be employed. For instance, the bank may send a one-time code via SMS to the user before completing a fiat transfer, mirroring common real-world security measures and preventing fraudulent transactions. 
\section{Evaluation}
\label{sec:evaluation}

\noindent\textbf{Prototype implementation.}
To evaluate the feasibility and practicality of \sysname, we implement a prototype demonstrating that its privacy-preserving design incurs minimal overhead compared to standard exchange platforms. The prototype comprises three main subsystems:
(i) a user registration system involving a platform, SSA, and users, where users submit PII and obtain VCs;
(ii) an exchange system where users anonymously present VCs and ZKPs to perform fiat-to-cryptocurrency transactions through banks and a Bitcoin test network; and
(iii) an audit system that enables regulatory authorities to de-anonymize users who misreport taxable gains or losses.
In addition, we implement a \textit{baseline} fiat-to-cryptocurrency exchange platform on the same server, simulating traditional exchange processes without privacy protections. Since \sysname does not alter the core exchange workflow, the primary differences lie in the added cryptographic operations for VC issuance, presentation, and ZKP generation.
All servers are deployed on Amazon EC2 instances (\texttt{c5.large}, Ubuntu 22.04), and the user client runs locally on a workstation with an Intel Xeon E5-2620 v4 (32-core, 2.10 GHz), 128 GB RAM, and Ubuntu 22.04.4. 
We also deploy our client on a Google Pixel 8a (Tensor G3 processor, 8GB RAM, Android 14) to evaluate mobile performance under resource-constrained environments.

Our implementation uses Python, with cryptographic primitives implemented via \texttt{Hyperledger AnonCreds} \cite{anoncreds-rs}, adhering to the Verifiable Credentials W3C Standard for credential issuance and presentation, and 2048-bit ElGamal encryption \cite{elgamal1985public} for identity protection, following NIST SP 800-57 guidelines \cite{nist80057}. We simulate the public ledger required for VCs using \texttt{Hyperledger Indy} \cite{indy}, and test cryptocurrency transfers on a local Bitcoin testnet \cite{lishaoyu1,lishaoyu2,lian2025facing}. Both the \textit{baseline} and \sysname use TLS secure connections for communication. However, the \textit{baseline} does not employ additional cryptographic techniques for privacy protection.

\begin{table}[ht]
\centering
\caption{Execution time across exchange phases}
\label{tab:exchange_process}
\renewcommand{\arraystretch}{1.03} 
\begin{tabularx}{\linewidth}{>{\raggedleft\scriptsize\arraybackslash}X|>{\centering\arraybackslash}X|>{\centering\arraybackslash}X|>{\centering\arraybackslash}X}
\hline
 & \textbf{Baseline} & \textbf{\sysname (PC)} & \textbf{\sysname (Mobile)} \\
\hline
\text{Identity Verification} & 2.63 ms & 464.87 ms & 3.40 s \\
\text{Bank Interaction} & 2.45 ms & 363.27 ms & 298.32 ms \\
\text{Bank Transfer} & 0.91 ms & 0.89 ms & 0.91 ms \\ 
\text{Bitcoin Transfer} & 170.01 ms & 170.67 ms & 169.26 ms \\ 
\hline
\end{tabularx}
\caption*{\scriptsize{The \textit{baseline} captures exchange logic without privacy protection. Identity verification includes plaintext account matching; bank interaction measures direct data transmission; bank transfer simulates VISA latency. Bitcoin transfer reflects the latency involved in transaction generation and network propagation on the testnet.
}}
\vspace{-0.1cm}
\end{table}

\noindent\textbf{Evaluation.}
We measured the time consumption of the three phases in \textit{baseline} and \sysname platforms to provide an intuitive view of the overhead introduced by our proposed privacy-preserving method. Additionally, for the mobile client, we also measure the average CPU and memory usage during the registration and exchange phases to evaluate resource consumption under constrained environments.

\begin{figure}[t]
\centering
    \subfigure[User registration]{
    \label{subfig:eva_user_registration}
    \includegraphics[width=0.225\textwidth]{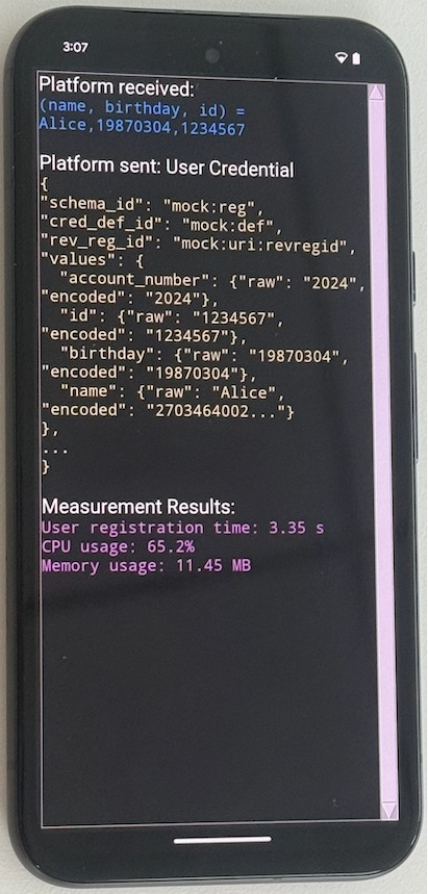}}
    \subfigure[Fiat-to-cryptocurrency exchange]{
    \label{subfig:eva_fiat_crypto_exchange}
    \includegraphics[width=0.2352\textwidth]{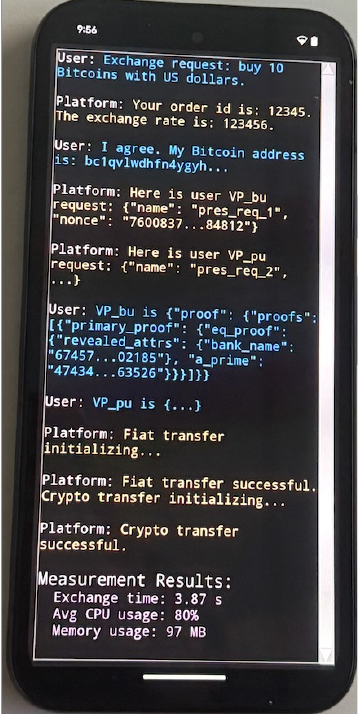}}
    \vspace{-7pt}
\caption{Mobile performance during registration and exchange} \label{fig:eva_mobile}
\vspace{-18pt}
\end{figure}

\noindent\textit{User Registration.} 
This process involves the user submitting PII, the platform verifying identity via the SSA API, the user retrieving schemas and credential definitions from the public ledger, and the platform issuing credentials to the user.
Since registration is a one-time operation performed before the exchange, it does not affect exchange latency.
We measure the time from the user's registration request to the receipt of the VC, excluding one-time system initialization steps, such as schema and credential definition generation and their upload to the public ledger.
Our results show that the \textit{baseline} requires 5.11 ms, while \sysname incurs 156.95 ms. On the mobile client, as shown in Figure~\ref{subfig:eva_user_registration}, the registration phase takes 3.35 seconds, with an average CPU usage of 65.2\% and memory consumption of 11.45 MB. These results demonstrate that the registration overhead introduced by \sysname remains within acceptable bounds and meets performance expectations across both desktop and mobile clients.

\noindent\textit{Fiat-to-cryptocurrency Exchange.} 
We also measured the time required for the exchange process between a user and the platform, as detailed in Table \ref{tab:exchange_process}. During identity verification, the \textit{baseline} system interacts with the user to verify the account number and password, whereas our system involves interactions for generating and verifying the user's VPs and ZKPs. During bank interaction, the \textit{baseline} transmits transaction data in plaintext, whereas \sysname requires the bank to decrypt ciphertexts and verify associated ZKP. Bank transfer latency is simulated under VISA-like conditions for both the \textit{baseline} and our system. 
Bitcoin transfer time is measured from transaction initialization to its first received by peers on the testnet. In addition, the mobile version of \sysname(as shown in Figure~\ref{subfig:eva_fiat_crypto_exchange}) exhibits an average CPU utilization of 80\% and memory usage of 97 MB during the exchange phase, with VP and ZKP generation taking 2.50 seconds. These results confirm that \sysname’s cryptographic operations—namely VP and ZKP generation and transmission—incur manageable overhead and do not pose a bottleneck in the exchange process, even under mobile constraints.

\noindent\textit{Auditing.} We evaluate the auditing process by measuring the time cost for the whole process for \sysname and \textit{baseline}. \sysname audit process involves interactions among three parties: the user, the platform, and the auditing authority, while the \textit{baseline} model involves only the platform and the auditing authority. \sysname process includes several steps: the user encrypts its SSN, generates a presentation and ZKP based on its VC, and sends all the information to the platform. The platform then verifies the user’s VP and ZKP, executes the order, and forwards the ciphertext along with the transaction record to the authority, which then decrypts the ciphertext to obtain the SSN. In contrast, the \textit{baseline} model involves the platform sending transaction details along with the user's SSN in plaintext to the authority. 
The total time consumption for the \sysname process is 362.52 ms, compared to 2.65 ms for the \textit{baseline}.

\section{Discussion and Conclusion}
\label{Conclusion}

We acknowledge that fiat payment infrastructures vary across jurisdictions. While \sysname targets the U.S. model based on platform-driven bank transfers, other regions may rely on different mechanisms such as manual bank transfers. Nevertheless, the employed VCs and ZKPs are general and can be adapted to alternative infrastructures through suitable protocol refinements. Moreover, although \sysname is designed to meet KYC and tax auditing requirements, supporting additional regulatory frameworks may require further extensions. Enabling such compliance while preserving user privacy remains a promising direction for future work.

In conclusion, we introduce \sysname, a privacy-preserving and regulation-compliant exchange system between fiat currency and cryptocurrency. The system ensures user anonymity in cryptocurrency transactions by concealing users' PII from the platform during the exchange via VC and ZKP techniques, preventing linkage between users' real-world identities and their cryptocurrency addresses. \sysname also enables the lawful de-anonymization of users for KYC compliance and auditing purposes, where authorities can validate and retrieve users' identities through cryptographic techniques. The implementation of \sysname demonstrates its practicality through the comprehensive evaluation.

\section*{ACKNOWLEDGMENT}
This work was supported in part by the US National Science Foundation under grants 2154929, 2247560, 2247561, 2331936, 2433904, and 2433905, and the Virginia Commonwealth Cyber Initiative (CCI). 

\bibliographystyle{acm}
\bibliography{main}

\end{document}